# SCHEME FOR QUANTITATIVE DESCRIPTION OF LONGITUDINAL DRIFTS IN THE FERMILAB LINAC AND THEIR CORRECTION

S. Rego[*1,2], R. Sharankova[†1], A. Shemyakin[‡1]
[1] Fermi National Accelerator Laboratory, Batavia, IL, USA
[2] École polytechnique, Palaiseau, France

*Abstract*

The Fermilab Linac accepts the 0.75 MeV H- ions from the front end and accelerates them to 400 MeV for injection into the Booster. Day-to-day drifts of the longitudinal trajectory in the Linac, reconstructed from phase readings of Beam Position Monitors, are at the level of several degrees. They are believed to cause additional losses both in the Linac and Booster, and are addressed by empirically adjusting the phases of Linac cavities. This work explores the option of expressing these drifts in terms of phase shifts in two cavities at the low-energy part of the Linac. Such a description allows for a simplified visual representation of the drifts, suggests a clear algorithm for their compensation, and provides a tool for estimating efficiency of such compensation.

## INTRODUCTION

*The Fermilab Linac*

The Fermilab Linac [1] consists of a Drift Tube Linac (DTL), a transition section, and a Side Coupled Linac (SCL). A magnetron ion source produces a pulsed 35 keV H- beam, which is transported in the Low Energy Beam Transport (LEBT) line to the Radiofrequency Quadrupole (RFQ). The beam accelerated to 750 keV in the RFQ travels through the Medium Energy Beam Transport (MEBT) and enters the Linac.

An overview of the DTL and SCL is shown in Figure 1. The DTL consists of 207 drift tubes spread across 5 "tanks" and accelerates the beam to 116.5 MeV. The SCL has 448 cells grouped into 7 "modules" that accelerate the beam to 401.5 MeV. Since the sections operate at different frequencies (201.25 MHz and 805 MHz respectively), there is a transition section in between consisting of a bunching and vernier cavities for longitudinal matching between the sections.

During regular operation, the Linac outputs about 25 mA of beam current in 35 $\mu$s long pulses at 15 Hz repetition rate with transmission efficiency $\geq 92\%$.

*Drifts in the Linac*

The beam centroid position in the Linac is reported by 33 Beam Position Monitors (BPMs). The BPMs are 4-button electrostatic pickups that provide information about the beam transverse position as well as its longitudinal phase with respect to the cavity RF reference line for every beam pulse. While the DTL has only 5 operational BPMs, the SCL is well instrumented with 3-4 BPMs per module. The BPM rms phase noise as reported in [2] is 0.1° of 201.25 MHz.

The absolute value of the BPM phase is not calibrated and depends on cable length. However, these offsets are believed to be stable and, therefore, relative changes in BPM phases are meaningful, reflecting actual changes in beam arrival times.

As an example, the changes of BPM phases over a week are shown in Figure 6b, with the scatter of several degrees. Such changes can cause additional losses in both Linac and Booster, and phases of several cavities are empirically adjusted in daily accelerator tuning to address the loss increase.

The origin of the phase drifts has not been definitively identified. Some likely factors include fluctuations in the ambient temperature and humidity that affect cavity resonance frequencies as well as phase and energy of the beam coming out of the RFQ. One of the difficulties in correcting such phase drifts is to find a simple way to numerically describe them. In attempt to address that, in this paper we present an approach based on fitting of the observed drifts to measured responses of the BPMs to varying of the tank phases.

*Paper layout*

The layout of the paper is as follows. The first section outlines the proposed description and correction scheme. Next, we illustrate the idea using a simple toy model of a sinusoidal motion. The following section describes the method of measuring the responses of BPMs to variation of tank phases. In the section after that, the drifts in two separate weeks taken from the historical data log are analyzed. Finally, we discuss the results and conclude.

## PROPOSAL FOR DRIFT DESCRIPTION AND CORRECTION

In single-particle tracking, the longitudinal motion is described in terms of deviations from motion of a synchronous particle. For the case of a linear accelerator, the synchronous particle is usually assigned to the motion of the centroid of the beam accelerated in the most optimal conditions.

Presently, the Linac doesn't have tools to deduce the beam absolute energy and phase in its intermediate points to follow the same procedure. Since the BPMs report an average beam phase per beam pulse, in this paper, we deal only with changes of BPM phases between different pulses and do not investigate phase stability withing a single pulse.

[*] sheldon.rego@polytechnique.edu
[†] rshara01@fnal.gov
[‡] shemyakin@fnal.gov

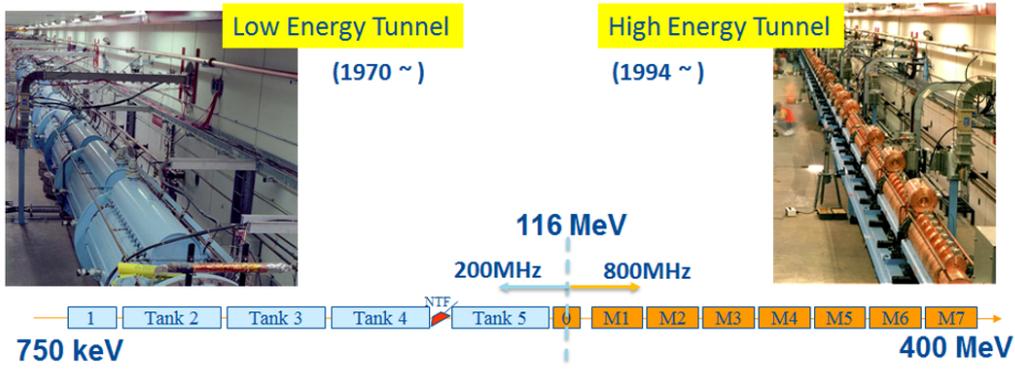

Figure 1: Drift Tube Linac and Side Coupled Linac module design

One of the sets of all BPM phases recorded for the same pulse (or an average of multiple pulses) is chosen as a primary or reference set that is an analog of the synchronous particle motion. Readings of BPM phases in other pulses are always presented as a set of differences, for each BPM, between the phase reading in this occurrence and in the reference one. Such set of phase differences will be referred to as "longitudinal trajectory". In this terms, the final goal is to stabilize the longitudinal trajectory, so its root mean squared (RMS) metric would be minimal.

The proposal is based on several idealized assumptions:

1. The BPM intrinsic electronic noise is small in comparison with beam-related drifts in BPM phases.
2. The phases and amplitudes of cavities in the SCL are significantly more stable than in the DTL and the front end, hence the drifts are originated primarily upstream of the SCL.
3. Within the typical drift amplitudes (order of 1° to 2°), the longitudinal dynamics can be described in linear terms.

Since in this paper we are interested only in motion of the beam centroid, the longitudinal trajectory is equivalent to the motion of a particle with energy and phase deviated from the synchronous particle's, projected to the phase axis. Such motion is described by a a second order differential equation (see, for example, [3], Eq.(2.41)). Any solution of this equation is uniquely defined by two parameters, which are typically chosen as initial energy and phase offset, $E_0$ and $\phi_0$. In the case of small perturbations, the phase offset $P_i$ measured by a BPM at the location of $z_i$ can be linearized,

$$P_i = \frac{\partial P_i}{\partial E_0} E_0 + \frac{\partial P_i}{\partial \phi_0} \phi_0 + \ldots . \quad (1)$$

Therefore, any longitudinal trajectory can be expressed as a linear sum of two "reference trajectories" proportional to the response for perturbation by the energy only $\{\frac{\partial P_i}{\partial E_0}\}$ and by phase offset only $\{\frac{\partial P_i}{\partial \phi_0}\}$. While in practice it is difficult to produce a pure energy or phase change, one can choose different, easily measurable reference trajectories that are combinations of perturbations in both energy and phase. This work considers motion only in a part of the accelerator (SCL), and the reference trajectory can be produced by varying the RF phase $\psi_j$ of one the tanks ($j$), which are, as mentioned before, the drift tube (Alvarez) cavities making up the DTL. In this case, any longitudinal trajectory $\{P_i\}$ can be presented as a linear combination of BPM phase responses to variation of RF phases in two tanks ($j_1$ and $j_2$):

$$P_i = \frac{\partial P_i}{\partial \psi_{j_1}} \delta \psi_{j_1} + \frac{\partial P_i}{\partial \psi_{j_2}} \delta \psi_{j_2} \quad (2)$$

The set of responses $\{\frac{\partial P_i}{\partial \psi_j}\}$ can be called the response matrix analogously to its use in the analysis of transverse motion [4]. When the response matrix is measured, description of any longitudinal trajectory can be reduced to just two numbers $\delta \psi_{j_1}$ and $\delta \psi_{j_2}$. Therefore, a drift of BPM phase readings in time can be presented as motion in ($\delta \psi_{j_1}, \delta \psi_{j_2}$) plane.

For the given choice of $\{\frac{\partial P_i}{\partial \psi_{j_1}}\}$ and $\{\frac{\partial P_i}{\partial \psi_{j_2}}\}$, calculation of $\delta \psi_{j_1}$ and $\delta \psi_{j_2}$ for a longitudinal trajectory requires readings in only two BPMs.

In absence of BPM noise and changes in fields of the cavities in the SCL between the time of measuring the response matrix and the drift, any pair of BPMs should give the same answer. To understand the effect of the noise, one can record the response matrix with respect to multiple tanks within a short period of time to avoid drifts, choose two of them as the reference, fit the rest to the reference using different BPM pairs, and check the scatter of ($\delta \psi_{j_1}, \delta \psi_{j_2}$) between different pairs.

An excessive scatter between the results from different BPM pairs in the analysis of drifts would indicate that the assumption of the stable longitudinal focusing in SCL is incorrect. As long as this scatter is significantly lower than deviation of the average from zero, actual application of the found tank phase shifts (with the opposite signs) should compensate the drift.

Before proceeding with the program explained above, it is useful to demonstrate the concept with a simple toy model.

## THE TOY MODEL

*Sinusoidal model*

To illustrate the proposal, let's consider purely sinusoidal motion,

$$P(z) = A \sin(kz + \phi_0), \quad (3)$$

producing a set of BPM readings $\{P_{ij}\}$ in locations $\{z_j\}$ in response to a change in RF phase at cavity $\{i\}$.

To model the reference trajectories $P_1$ and $P_2$, the phases of two cavities are varied. The cavities are located upstream of the section under consideration and separated from the beginning of that section by the phase advances $\phi_1$ and $\phi_2$. If the cavities are optically thin, the cavity phase variation changes only the energy. In terms of the longitudinal motion, it corresponds to jumps of the phase derivative ("kicks") $K_1$ and $K_2$:

$$\left.\frac{dP}{dz}\right|_{-\frac{\phi_i}{k}} = K_i, \quad (4)$$

where $i = 1, 2$. The coefficients $K_i$ correspond to basis trajectories $P_i$ (downstream of the kick) defined as,

$$P_i(z) = \frac{1}{k} \sin(kz + \phi_i), \quad (5)$$

in such a way that any arbitrary trajectory (Equation 3) can be decomposed in an equation of the form,

$$P(z) = K_1 P_1(z) + K_2 P_2(z). \quad (6)$$

Based on this definition, the expression for $K_i$ is found to be,

$$\begin{aligned} K_1 &= Ak \frac{\sin(\phi_2 - \phi_0)}{\sin(\phi_2 - \phi_1)}, \\ K_2 &= -Ak \frac{\sin(\phi_1 - \phi_0)}{\sin(\phi_2 - \phi_1)}. \end{aligned} \quad (7)$$

Obviously, this transformation works only if the kicks are distinguishable, ie. $\phi_2 - \phi_1$ is not an integer multiple of $\pi$. When $|\phi_2 - \phi_1| = \frac{\pi}{2}$, the coefficients $K_1$ and $K_2$ are of the smallest amplitude,

$$\begin{aligned} K_1 &= Ak \sin(\phi_2 - \phi_0), \\ K_2 &= -Ak \sin(\phi_1 - \phi_0). \end{aligned} \quad (8)$$

*Numerical demonstration and pair predictions*

In this section, the model above is first demonstrated numerically with test numbers, then the idea of BPM pair predictions is illustrated.

Let the two basis trajectories be $P_1(z) = \sin(z/3)$ and $P_2(z) = \sin(z/3 + \pi/2)$ and choose the trajectory to be described as $P(z) = 1.5 \cos(z/3 + 0.3\pi)$. Gaussian noise with amplitude 0.02 and unit standard deviation is added to $P(z)$. 40 points are used in the fit to keep the difference between successive points dominated by signal rather than noise. In the actual data, these 40 points would represent individual BPM readings, collectively providing information on the longitudinal trajectory of the particle.

Instead of fitting the entire trajectory to the reference trajectories (a "global fit"), one can fit two consecutive points at a time, generating for each pair its values of $K_1$ and $K_2$. This procedure is referred to as "pair predictions" or "BPM pair predictions" when the points represent BPM readings in the real data. Figure 2a shows the global fit and Figure 2c shows the pair prediction spread. Note than when the noise is not added, all the fit coefficients in Figure 2c fall on the same point. The scatter of $(K_1, K_2)$ values gives, therefore, a representation of the fit precision.

When the reference trajectories become closer to each other and farther from being separated by $\pi/2$, the value of $\sin(\phi_2 - \phi_1)$ decreases, and, correspondingly, the amplitudes $K_1$ and $K_2$ of corrections increase. Note that the errors of reconstruction $\delta K_1$ and $\delta K_2$ defined by the noise increase as well, and the residuals of the corrected trajectory stay the same. However, in a practical application the increase of the correction coefficients breaks the assumption of linearity of responses. For the case when the reference trajectories (and corrections of a drift) are induced by shifting of cavity phases, the resulting kick by the downstream cavity becomes dependent on the kick by the upstream one. Therefore, the optimum choice of the reference trajectories is important. For the case of the toy model, it corresponds, as mentioned above, to

$$\phi_2 - \phi_1 = \frac{\pi}{2}. \quad (9)$$

The analog of this requirement in a real lattice can be orthogonality of the basis vectors in the sense of minimizing the cosine of the angle $\alpha$ between them calculated as

$$\cos \alpha = \frac{\langle P_{1j}, P_{2j} \rangle}{|P_{2j}||P_{1j}|}, \quad (10)$$

where the angle brackets represent the inner product of two vectors.

Note that numerically the angle defined by Eq. (10) becomes dependent on the specific choice of measurement points. In the toy model with equidistant measuring points, Eq.(10) produces exactly the condition of Eq.(9) only if these point cover an integer number of half-periods. However, the deviations at other choices are not dramatic, and the the procedure of choosing the reference trajectories with minimizing the expression in Eq.(10) should work reasonably well.

An important assumption of the proposal is that the optics of the section where the trajectory is measured is stable. If, for example, one cavity drifts, it introduces an energy change that appears in the trajectory as a jump of the phase derivative. Figure 2b helps to visualize how it may appear in real data. The green curve emulating the measurement experiences a sudden change of its derivative, and the global fit (yellow curve) is no longer good. In pair predictions (Fig. 2d), the cloud of points is split into two: the top from the first half of pairs and the bottom for the second half after the derivative jump. The global fit lands in the middle along with the average of the cloud.

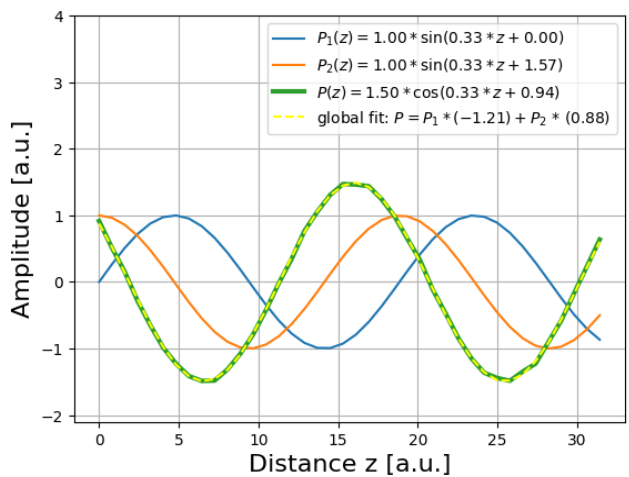

(a)

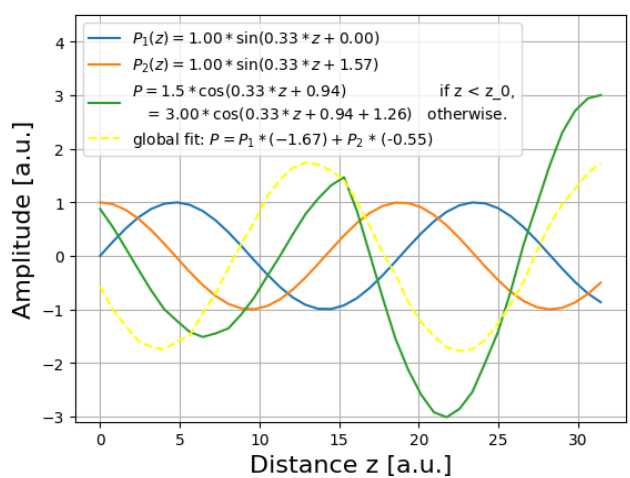

(b)

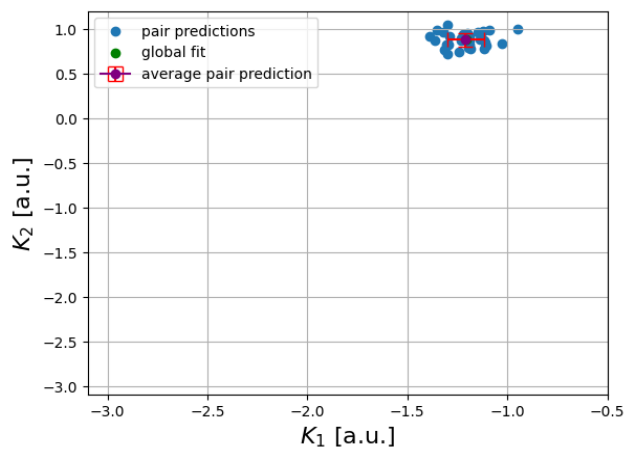

(c)

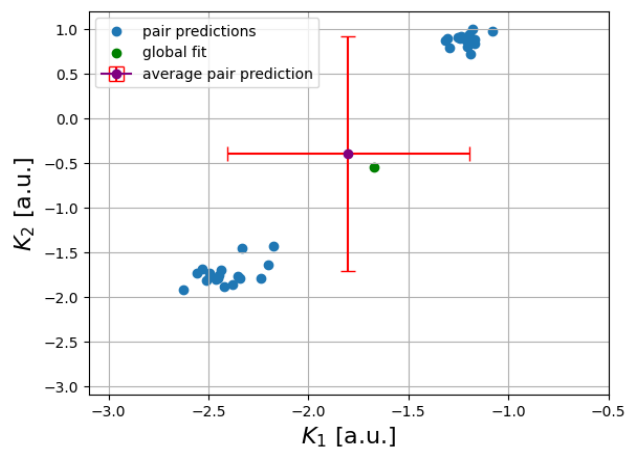

(d)

Figure 2: Pair predictions toy model (a) & (b) without phase kick (c) & (d) with phase kick

A similar grouping of pair predictions in measured data would indicate a specific cavity drifting significantly more than others.

## RESPONSE MATRIX

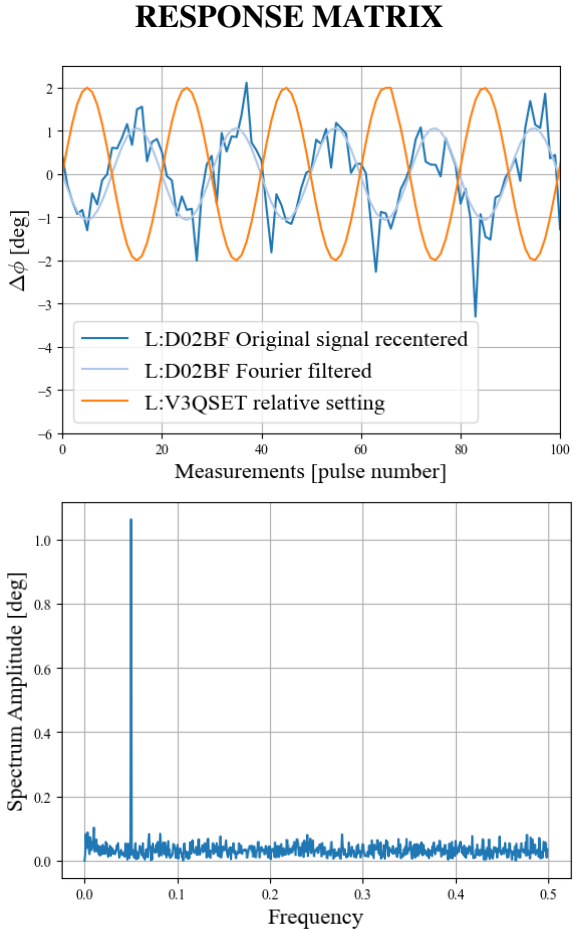

Figure 3: Response of the L:D02BF BPM phase to oscillation of the Tank 3. Top: beginning of oscillations. Orange - the driving signal, dark blue - the BPM signal, light blue - sine curve corresponding to the Fourier component at the driving frequency. Bottom: Fourier spectrum of the BPM signal for 75 periods, normalized to show the amplitude of each Fourier component.

### Recording the BPM responses

The first step to follow the proposal is to choose the reference trajectories. For that, the matrix of responses of BPM phases to variations of phases of five tanks in the DTL $T_j$, $j = 1...5$ and the MEBT bunching cavity $T_0$ is recorded. For brevity, in the rest of the paper, the MEBT bunching cavity is included into the collective "tanks" in most of cases.

To improve the measurement accuracy while still staying in the range of small perturbations, the measurements were performed by oscillating the tank phases and analyzing the BPM response at the driving frequency following Ref. [5]. With proper normalization, the amplitude of this response divided by the amplitude of the driving signal provides the response matrix $\frac{\partial P_i}{\partial T_j}$.

The response matrix was recorded during the Linac study period on July 19, 2023. For data collection, each $T_j$ phase was sinusoidally oscillated, one at a time, with the amplitude of 2° around its default value, changing the phase setting after each beam pulse. For the study, the average beam pulse frequency was 4.3 Hz, and one oscillation period was 19.93 pulses corresponding to the oscillation frequency of 0.215 Hz. The total number of recorded periods was 75 ($T_2$ and $T_3$) or 45 (all the others). Signals of all available 28 BPMs in the HE Linac were saved for each measurements. However, three of them are found noisy and are not used in the analysis.

An example of the signal recorded by one of BPMs in the Transition section, referred to as L:D02BF and its Fourier spectrum are shown in Figure 3. Also plotted in the top figure is the driving signal for the cavity phase of cavity $T_3$, L:V3QSET. The frequency on the horizontal axis of the bottom figure is expressed as the number of the corresponding component in the Fourier spectrum divided by the total number of the measured points. For technical reasons, the beam pulses went not always at even intervals. However, it does not affect the analysis since the cavity phase was changed every time only after recording the BPMs in the previous pulse. To estimate the frequency of the component in Hz, one can multiply the number on the horizontal by the average pulse frequency, 4.3 Hz.

The resulting longitudinal trajectories, equivalent within the normalization by the driving signal to the response matrix, are shown in Fig.4a. The error bars are calculated as $\frac{\sigma_n}{\sqrt{2}}$ [5], where $\sigma_n$ is the rms value of the Fourier components surrounding the driving frequency peak.

Note that while the amplitude of the driving signal was the same for all cases (2°), the amplitudes of the resulting longitudinal trajectories differ significantly.

### Choice of the reference trajectories

The cosines calculated with the data in Figure 4a are shown in Table 1. The responses to Tank 3 ($T_3$) and Tank 4 ($T_4$) are chosen as the reference trajectories as cosine of the angle between them is minimal.

### Consistency check

If the measurement noise and drifts between measurements are negligible, each longitudinal trajectory recorded in the response matrix measurements should be a linear combination of the two reference ones, with the coefficients correspond-

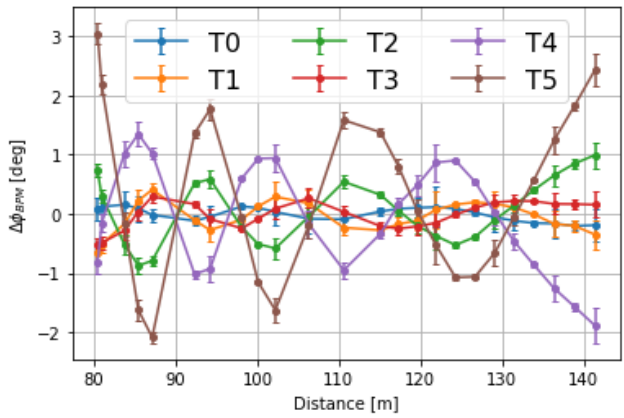

(a)

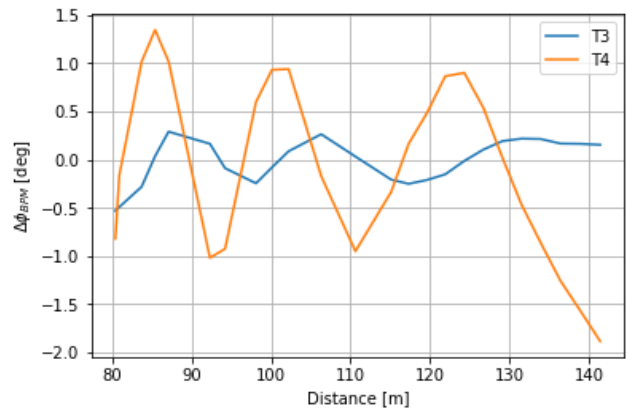

(b)

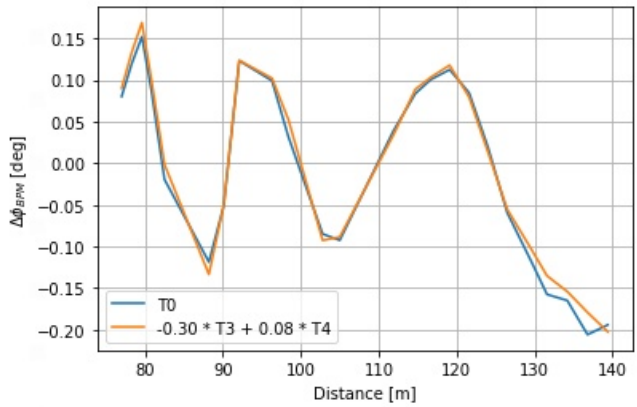

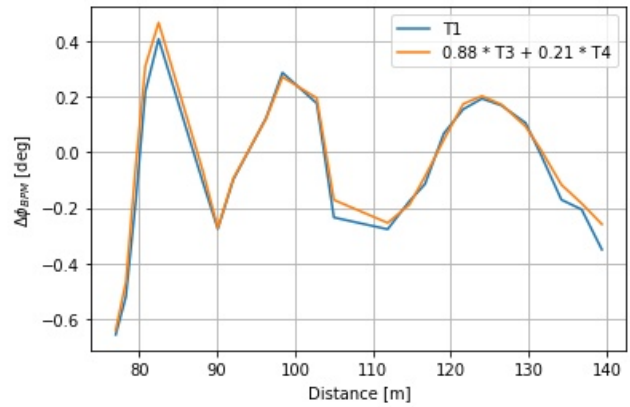

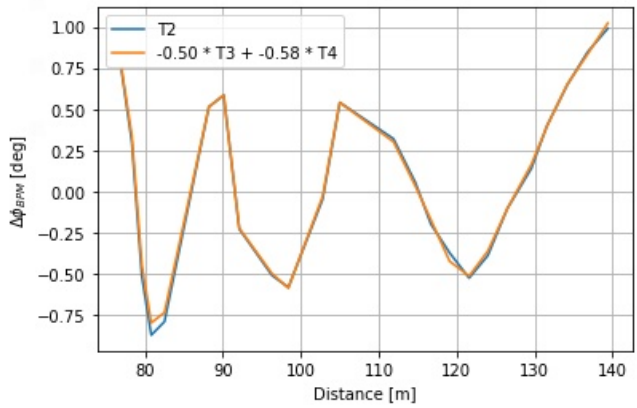

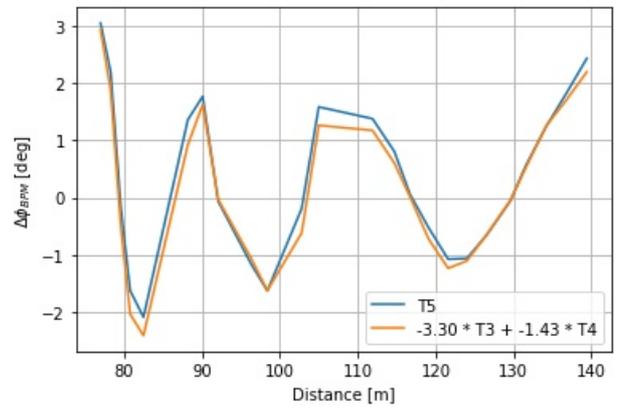

(c)

Figure 4: Measured responses for (a) all tanks (b) $T_3/T_4$ (c) reduction to $(T_3, T_4)$ basis

| Basis choice | $\cos \alpha$ |
|---|---|
| $(T_0, T_1)$ | 0.55141 |
| $(T_0, T_2)$ | 0.02089 |
| $(T_0, T_3)$ | 0.32038 |
| $(T_0, T_4)$ | 0.56554 |
| $(T_0, T_5)$ | -0.0513 |
| $(T_1, T_2)$ | -0.3349 |
| $(T_1, T_3)$ | 0.78995 |
| $(T_1, T_4)$ | 0.60337 |
| $(T_1, T_5)$ | -0.6469 |
| $(T_2, T_3)$ | 0.21590 |
| $(T_2, T_4)$ | -0.8082 |
| $(T_2, T_5)$ | 0.89384 |
| $(T_3, T_4)$ | 0.01287 |
| $(T_3, T_5)$ | -0.1818 |
| $(T_4, T_5)$ | -0.7779 |

Table 1: Cosine of the angle between possible basis choices. Calculated as $\langle T_x, T_y \rangle / \sqrt{\langle T_x, T_x \rangle \langle T_y, T_y \rangle}$ where $\langle \cdot, \cdot \rangle$ is the inner (dot) product.

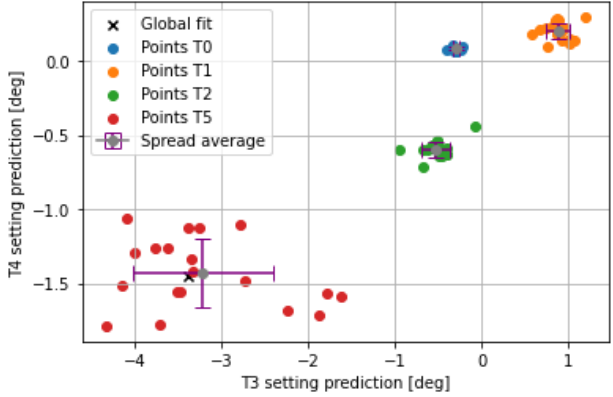

Figure 5: BPM pair predictions for $T_0, T_1, T_2, T_5$

ing to the changes in $T_3$ and $T_4$ phases that create the same trajectory. Results or such fitting are shown in Figure 4c.

A possible way to quantify resolution of $T_3$ and $T_4$ prediction is to perform their reconstruction from pairs of BPMs. The result is shown in Figure 5. In addition, the averages of each group with the error bars representing their standard deviations are plotted as well. The resolution of the overall prediction can be estimated as the standard deviation divided by the square of the number of the pairs. While the spread of individual points is significant, particularly for the $T_5$ case, the overall resolution stays sub-degree while the total shift is at several degrees. Such resolution should allow to correct the majority of a drift.

## ANALYSIS OF DRIFTS

The next part of the analysis is to apply the developed method to historical data, quantify the drifts, and estimate how well they could be compensated.

Two periods are analyzed, June 11-19, 2023 and July 3-13, 2023. In the text below these periods are referred to as June and July data, respectively. The archived BPM data are retrieved for each day between 11PM and midnight, and the data are averaged over this hour. The average of the beam longitudinal trajectories in each data set differed significantly from the trajectory saved on July 19, 2023, when the reference trajectories were recorded, with maximum deviations above 5°, requiring large corrections that are outside of the linear model. Therefore, a more modest goal is pursued below, limiting the analysis to drifts within each week. Specifically, to estimate weekly drifts, the values for first day of each time period are subtracted for the data for other days. The results are shown in Figures 6a & 6b.

Similarly to the previous section, data for BPM pairs are translated into $(T_3, T_4)$ pairs using the reference trajectories recorded on July 19, 2023.

The results of this procedure are shown in Figures 6c & 6d. The averages of the clouds with error bars showing resolution of the prediction for the corrections are shown in Figure 6e & 6f. The precision for each group is calculated as the standard deviation divided by square root of the number of BPM pairs.

Since the prediction resolution is significantly smaller than the averages, the corrections to apply to the Tank 3 and Tank 4 phase are reasonably well defined. An example of a possible correction is shown in Figure 8. Applying of the correction procedure would allow to significantly decrease the BPM phase deviations.

### Estimation of correction efficiency

The severity of a drift (from the first day in the week) can be expressed as the standard deviation of all BPM readings for the given longitudinal trajectory, $\sigma_0$.

To simulate applying the correction to the historical data, a differential trajectory is calculated as the difference between the drift longitudinal trajectory and the one reconstructed with the calculated correction values of $(T_3, T_4)$. If this differential trajectory has the standard deviation of $\sigma_{\text{corr}}$, the efficiency of the correction can be expressed as the ratio of $\frac{\sigma_0}{\sigma_{\text{corr}}}$. For the two analyzed periods, this ratio is shown in Figure 7. The drift amplitude is reduced by a factor of 1.2 to 4.

## DISCUSSION ON PRACTICAL IMPLEMENTATION

The results presented in the previous sections indicate that changes in the beam phases over a week fit well with a linear combination of two reference trajectories. On the other hand, the drifts in a month appear to be too large to be addressed with linear corrections. Since the time period of about one week was chosen somewhat arbitrarily, further research could be done to find an optimal time scale for corrections. Also, it is not clear yet what is the time scale

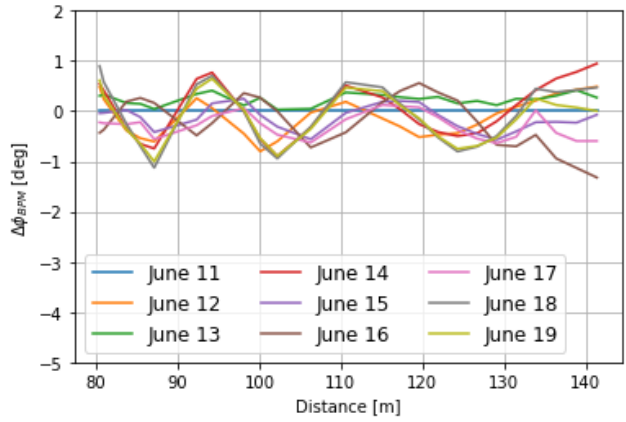
(a)
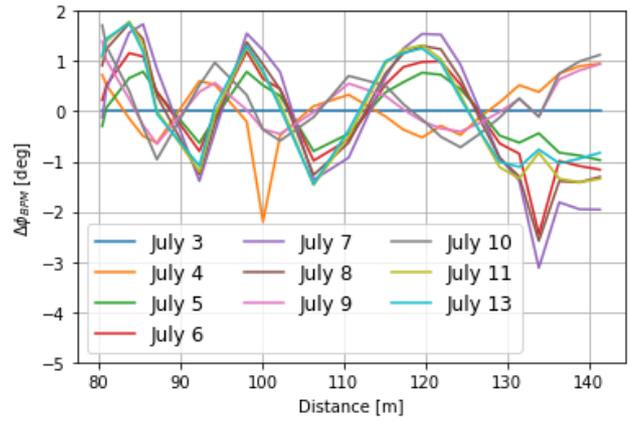
(b)
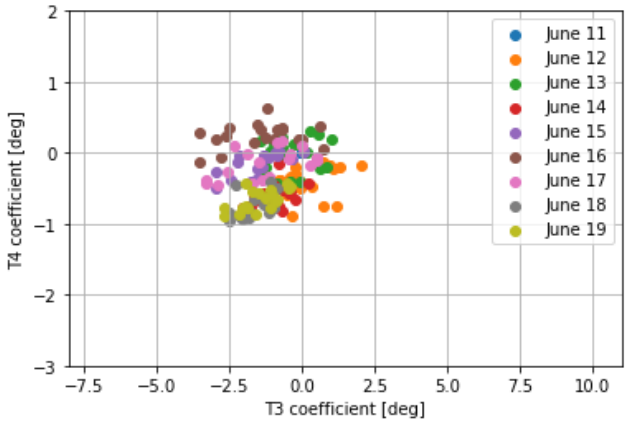
(c)
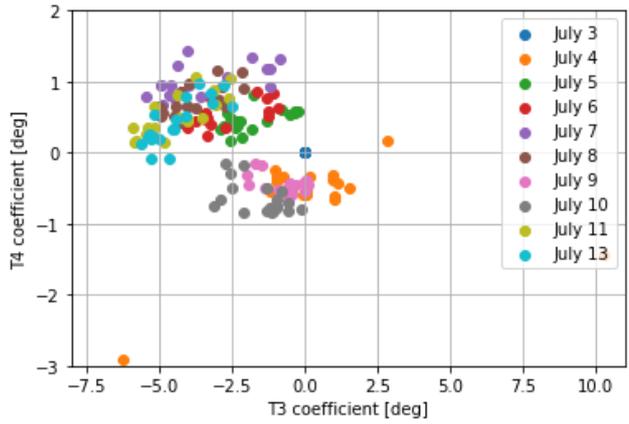
(d)
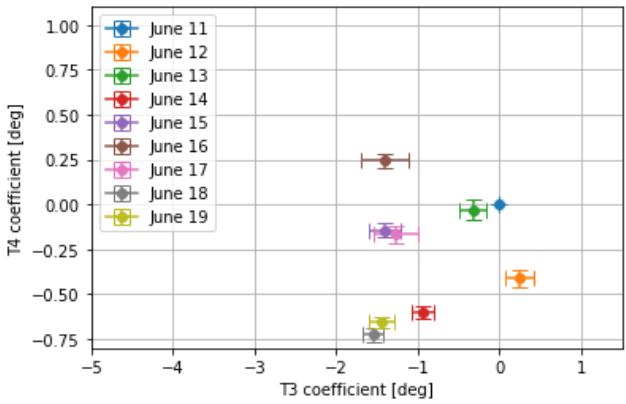
(e)
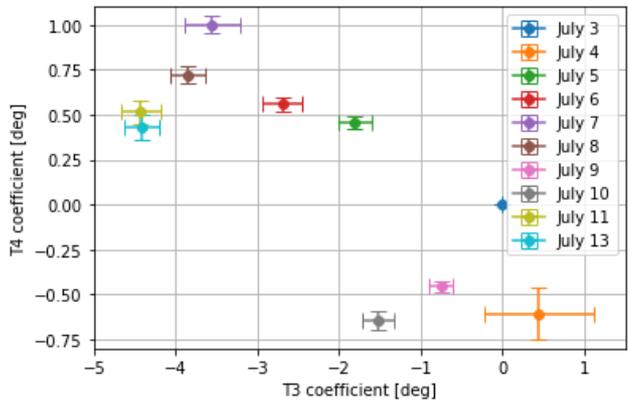
(f)

Figure 6: Historical data analysis (a) June drift data (b) July drift data (c) July pair predictions (d) July pair predictions (e) Averages of June predictions (f) Averages of July predictions. Error bars in (e) and (f) represent averages divided by $\sqrt{N}$.

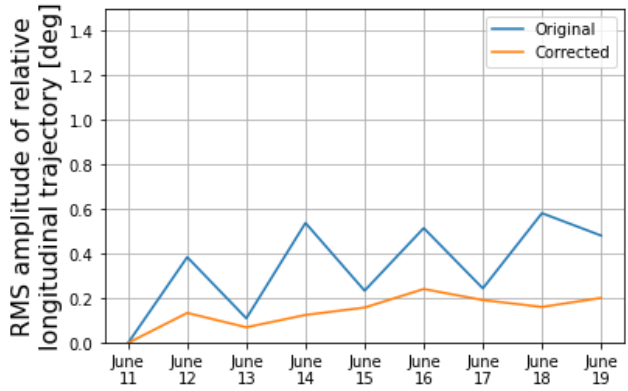
(a)

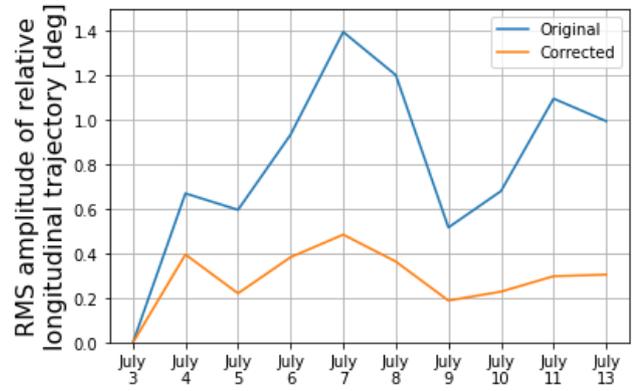
(b)

Figure 7: Calculated reduction in drifts for June '23 (a) and July '23 data sets.

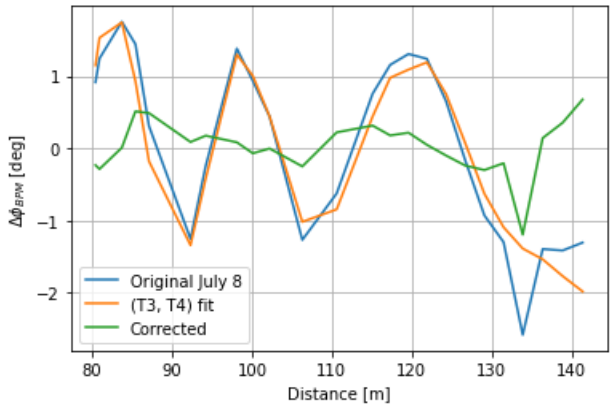

Figure 8: Best fit correction estimated on drift of July 8th relative to July 3rd

for the response matrix itself to drift significantly. Since both response matrix measurements and small corrections can be made without interference with regular operation, presently the following procedure (not automated yet) is being considered:

1. Generate response matrices for all DTL tanks
2. Select reference trajectories based on orthogonality
3. Keep monitoring the BPM phase stability
4. Apply correction if a significant drift from reference is observed; it can be made in steps so that the corrections are applied with some coefficient less than 1
5. Repeat at the beginning of the following correction period when response matrices for DTL tanks have shifted significantly

Since the drifts are measured with respect to the first day of the period, the correction must be implemented as changes to the absolute phase of the tanks with respect to start of the correction period, not with respect to the previous day. The correction strategy is expected to be tested on real-time data during standard beam operations.

## CONCLUSION

A procedure quantifying the longitudinal drifts in the Fermilab Linac and their corrections was developed. It consists of several steps: firstly, responses of BPM phases in the high-energy section of the Fermilab Linac to variation of phases of cavity sets (tanks) in its low-energy section are recorded. Responses to two tanks are chosen as reference trajectories according to maximum orthogonality between different trajectories. Next, deviations of BPM phases over time are fitted to a linear combination of the reference trajectories. Coefficients of this linear combination are equivalent to the changes of phases of two corresponding tanks to produce the same deviations. The values of these changes can be used as numerical characterization of the longitudinal trajectory drifts.

If the linear fit is applied to individual pairs of BPMs, the resulting scatter of the coefficients indicates the precision of fitting. If the precision of the coefficient reconstruction is significantly better than the observed drift expressed by the average coefficient values, these values can be used for the drift correction.

Applying this procedure to historical data shows that the drift can be reduced by a factor of 2 to 3.

## ACKNOWLEDGEMENTS

This manuscript has been authored by Fermi Research Alliance, LLC under Contract No. DE-AC02-07CH11359 with the U.S. Department of Energy, Office of Science, Office of High Energy Physics. The study was performed as part of the Helen Edwards Summer Internship Program 2023.